%% file: document.tex
\documentclass[lnbip,sechang,a4paper]{svmultln}

\usepackage{makeidx}
\usepackage{multicol}
\usepackage{mathtools}
\usepackage{cite}
\usepackage{setspace}
\usepackage[hyphens]{url}
\usepackage[hidelinks]{hyperref}
\usepackage{lipsum}
\usepackage{graphicx}
\usepackage{curve2e}
\usepackage{epstopdf}
\usepackage{subfigure}
\usepackage{listings}
\usepackage{floatflt}
\usepackage{multicol}
\usepackage[font=scriptsize]{caption}
\usepackage{comment}
\usepackage{todonotes}
\usepackage{paralist}
\usepackage{booktabs}
\usepackage[final]{changes}
\usepackage[bottom]{footmisc}

\begin{document}

\newcommand{\system}{OPlatform}
\newcommand{\obox}{OBox}
\newcommand{\ometer}{OMeter}
\newcommand{\nano}{NanOMeter}
\newcommand{\sysname}{\textsc{Energy Switch}}

\mainmatter

\title{Micro-accounting for optimizing and saving energy in smart buildings}

\author{Daniele	Sora\inst{1} \and
Massimo	Mecella	\inst{1} \and
Francesco	Leotta	\inst{1} \and\\
Leonardo Querzoni \inst{1} \and
Roberto Baldoni \inst{1} \and
Giuseppe	Bracone	\inst{2} \and\\
Daniele	Buonanno\inst{2} \and		
Mario	Caruso	\inst{2} \and
Adriano	Cerocchi\inst{2} \and		
Mariano	Leva\inst{2}}

\tocauthor{Daniele Sora et al.}
\authorrunning{Daniele Sora et al.}

\institute{
Sapienza Universit\`a di Roma, Italy\\
\texttt{\{sora,mecella,leotta,querzoni,baldoni\}@diag.uniroma1.it}
\and 
Over Technologies, Italy\\
\texttt{\{g.bracone,d.buonanno,m.caruso,a.cerocchi,m.leva\}@overtechnologies.com}
}

{\let\newpage\relax\maketitle}
	
\begin{abstract}
	Energy management, and in particular its optimization, is one of the hot trends in the current days, both at the enterprise level (optimization of whole corporate/government buildings) and single-citizens' homes. The current trend is to provide knowledge about the micro(scopic) energy consumption. This allows to save energy, but also to optimize the different energy sources (e.g., solar vs. traditional one) in case of a mixed architecture. In this work, after briefly introducing our specific platform for smart environments able to micro-account energy consumption of devices, we present two case studies of its utilization: energy saving in offices and smart switching among different energy sources. 
	
	\let\thefootnote\relax\footnote{\textbf{This is a post-peer-review, pre-copyedit version of the article published
		in Lecture Notes in Business Information Processing, vol 249. Springer,
		Cham. The final authenticated version is available online at: \url{https://
		doi.org/10.1007/978-3-319-39564-7_15}}
	}
\end{abstract}

\input{introduction.tex}
\input{architecture.tex}

\input{banks.tex}
\input{switch.tex}
\input{conclusions.tex}


\input{finalRefs.bbl}
\end{document}

%% file: introduction.tex
\section{Introduction}

Energy management, and in particular its optimization, is one of the hot trends in the current days, both at the enterprise level (optimization of whole corporate/government buildings) and single-citizens' homes. 
Energy efficiency is generally function of \emph{out-door} techniques -- renewable energy, smart energy production and distribution, etc. -- and \emph{in-door} techniques; in particular, very few energy managers -- each of us can be an energy manager of his own home -- can state ``who, when and why is consuming'', conversely this knowledge is fundamental in order to cut energy cost. 
As stated in \cite{ehrhardt2010advanced}, energy consumption feedback increases user's awareness, motivation and responsibility and can result in more than 12\% of energy saving when the feedback is real-time and at the granularity of the single electrical appliance.

Pervasive systems and Internet-of-Things (IoT) are gaining popularity due to their invisible integration into everyday life. In particular, among many other smart sensors and actuators,  users are starting to introduce and integrate Home/Building Energy Management Systems (H/BEMSs) in their environments. Such systems allow users to monitor, control and optimize energy consumption \added{\cite{carusoThesis}}.

Energy saving can be achieved in many ways: by optimizing appliance usage, by shifting energy consumption to off-peak hours or by dimming light brightness based on the available natural light \cite{rojchaya2009development}.
In \cite{ricciardi}, for example, authors present a system to save energy in PC networks by
turning PCs on only when really needed thanks to the WoL (Wake on LAN) technology.
An energy model to calculate achievable energy savings is proposed along with real world scenario evaluation resulting in remarkable savings in terms of both money and $CO_2$ emissions. 
The process of energy monitoring clearly involves, besides final users, also energy utility providers; with the term Advanced Metering Infrastructures (AMI), we refer to the bidirectional infrastructure and communication system at the base of smart grids \cite{hart2008using,masiello2010demand}.
Recent B/HEMS make use of social networks to motivate energy saving; in \cite{han2011green}, for example, energy data are collected through a Zig-Bee network and are then published in a social network to let the users figure out how efficient its home appliance is, compared to others.
The trend is therefore to provide knowledge about the micro(scopic) energy consumption. 

In this work we briefly present in Section \ref{sec:platform} our platform, named \system~\footnote{The ``O'' prefix assigned to the components (\ometer{}  and \obox{}) and to the platform itself (OPlatform) comes from the name of the spin-off, Over, which has engineered the research.}, for smart environments able to micro-account energy consumption of devices, at the level of single power line, which allows at the same time the actuation of devices, thus being also an energy-aware smart space automation solution. Then in Sections \ref{sec:banks} and \ref{sec:switch} we present two current applications of it: a case study of energy savings in offices and its usage for optimizing the smart switch-off among different energy sources. Finally, Section \ref{sec:end} concludes the paper.

%% file: architecture.tex
\section{The OPlatform}
\label{sec:platform}

\begin{figure}
\centering
\includegraphics[width=0.9\columnwidth]{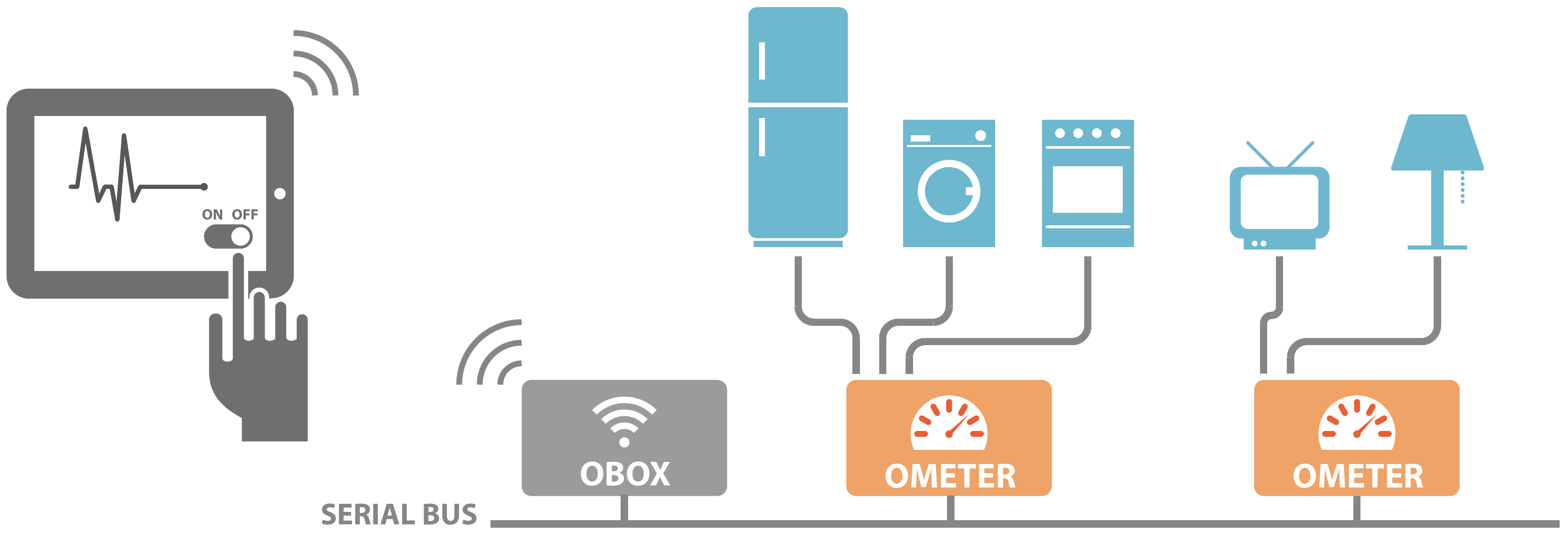}
\caption{The platform for energy micro-monitoring}
\label{fig:arch}
\end{figure}

The \system{} consists of two main devices: the \ometer{} and the \obox{}, which are responsible, respectively, for controlling electrical appliances and for accessing the whole space automation system through web based technologies. 
The typical deployment of the \system{} consists of a single \obox{} and several \ometer{}s, the number of which depends on the number of electrical devices to be controlled/monitored (Fig. \ref{fig:arch}).

\noindent \textbf{\ometer{} --} The role of an \ometer{} is to actuate electrical loads and to sense binary input commands. 
Each \ometer{} can handle up to 8 independent electrical circuits 
providing for their actuation by means of relays which bear a maximum current of 16 amperes. Each of these circuits can be, depending on the granularity that user wants to control, a power line to which many loads are connected to, or an individual electrical load. The \ometer{}, in addition, handles up to 16 dry contact inputs 
such as toggle switches, PIR (Passive InfraRed) sensors, magnetic contact sensors, etc. As each load can be switched on/off, the \ometer{} is able to monitor its power consumption in terms of voltage, current, power factor, apparent, active and reactive power.

As the system is designed to monitor power consumption and prevent its waste, in the design phase we payed special attention to the energy consumption of the \ometer{}s themselves; a system that consumes more energy  than it allows to save would be, in fact, useless. This is particularly critical considering that the system will be active 24/7.
For this reason magnetic latching relays,
have been adopted; this type of relay has the advantage that one coil consumes power only for an instant, while it is being switched, and the relay contacts retain this setting across a power outage. The base current drawn by the \ometer{} is 150 milliamps and the contribution due to the magnetization of the relay is negligible resulting in a total power absorption of less than 2W.

As far as the electrical consumption monitoring, an \ometer{} is able to read the power drained by any load (up to 16 A) in a  230V, 50Hz electric network. For this, the \ometer{} is equipped with a microprocessor~ %
that samples, at very high frequency, instantaneous power data of each output, thus allowing to calculate TRMS (True Root Mean Square) values of the alternate current. This is particularly important for measuring non-linear loads (such as induction furnaces, PCs, displays), whose current waveform is not sinusoidal but typically distorted.

An \ometer{} explicitly provides information about apparent power, current and power factor, other measurements, such as voltage, active power and reactive power can be indirectly calculated by the \obox{}.

On the one hand, power consumption values can be inquired at any time through an explicit request, on the other one, they can be spontaneously notified
 by the meter itself according to different modes: 
\begin{inparaenum}[(a)]
\item at regular (customizable) time intervals;
\item when the difference between the current value and the last sent value is greater than a customizable threshold;
\item when the difference between the current value and the last sent value is higher than a customizable percentage of the latter.
\end{inparaenum} The \ometer{} can be configured to work in one of the three above modes, the choice depends on the desired accuracy but should take in account also the bus load. 


Whereas some applications, such as load monitoring or anomaly detection in electrical appliances, require instantaneous readings, other ones, e.g.,  electric bill analysis, may require 
more coarse grained values. For this purpose, the \ometer{} continuously calculates and stores average values of all the measurements; when average measures are explicitly requested, it replies with current average values and then resets such values and restarts calculating them.

In summary, the \ometer{} offers: 

\begin{itemize}

\item three primitives for synchronously controlling outputs and inputs:
	\begin{itemize}
	\item \textit{controlOut}, given an output number and a value (on or off) it switches on or off the corresponding output channel;
	\item \textit{readOutState}, given an output number, it will reply with the current state of the output channel; 
	\item \textit{readInState}, given an input number it will reply with the current state of the input channel;
	\end{itemize}
\item two primitives for asynchronously \replaced{notifying}{notify} informations about outputs and inputs:
	\begin{itemize}
	\item \textit{inVariation} is triggered when an input channel changes its state and informs about its new state;
	\item \textit{outVariation} is triggered when an output channel changes its state and informs about its new state;
	\end{itemize}
\item two primitives for synchronously \replaced{retrieving}{retrieve} power measurements:
	\begin{itemize}
	\item \textit{readMeasures}, given an output channel it will reply with current apparent power P, power factor (PF) and current (C);
	\item \textit{readAvgMeasures}, given an output channel it will reply with the average values of apparent power P, power factor (PF) and current (C);
	\end{itemize}
\item one primitive for asynchronously notify informations about power measurements:
	\begin{itemize}
	\item \textit{powerVariation}, triggered when one of P, PF, and C changes its values according to the rules explained above.
	\end{itemize}
\end{itemize}

\ometer{}s represent the nodes of a distributed system, and are interconnected via a serial bus, whose transfer rate is 9600 baud, that carries data and power (12V) simultaneously. The low bandwidth allows longer deployments of the bus, as well as more reliable communication on noisy channels. The units are able to coordinate themselves autonomously, without the presence of a master unit, by exchanging appropriate messages defined in a specific communication protocol.  
A wired solution has been adopted, rather than a wireless one, to improve reliability in transferring data even though it requires a more expensive and intrusive installation. Wireless solutions can incur in interferences and connection problems when the number of installed appliances is high. Moreover, wireless solutions mainly rely on batteries, thus requiring a certain maintenance effort.

Each message sent on the bus, in addition to the payload, has an header identifying sender and recipient; once sent, the message is perceived and analyzed by all the units on the bus, but is processed only by the unit having the address specified in the recipient field of the message and discarded by the other units.

\noindent \textbf{\obox{} --} The \obox{} is an embedded PC that is physically connected to the bus via a RS232 serial port and to the building/home network, and it plays the role of bridge between the several \ometer{}s deployed in the environment and the Web clients of the system.
It runs a lightweight web server and offers RESTful interfaces to configure, control and monitor the meters.
The \obox{} is connected to the serial bus as well, it is then in effect a node of the system and operates in the same way of a \ometer{}, thereby generating messages that control the various units. It differs from the \ometer{}s in the sense that it can not perform any electrical actuation or sensing. In addition, unlike others \ometer{}s units, the \obox{} is interested in all messages passing on the bus and does not discard any, as it must be always up to date about the context of the devices deployed in the home. The average power absorption of an \obox{} is about 8W.  

The \obox{} plays both roles of controller and supervisor. Being a controller, it is able to send commands to devices, for example to turn on or off a light. At the same time it must be able to work as a supervisor, i.e., it is able to continuously monitor the states of all the actuators and the sensors deployed in the house in order to be always aware of the environmental context. On the one hand the \obox{} is able to communicate with the \ometer{}s through the messages defined in the bus protocol and implemented in a specific driver, therefore it is able to operate home devices and to monitor their state and energy consumption both synchronously and asynchronously.
On the other hand, it abstracts the bus protocol and offers high level functions to REST clients hiding all the technical details of the underlaying protocol. 
A Web browser-based user interface has been designed to make the system accessible from many different devices (PCs, smartphones, tablets) 
without installing any specific software. Despite the fact that the interaction with the system occurs via the browser, the \obox{} communicates with clients by sending raw data rather than formatted HTML and the client itself updates the UI by properly rendering such data. In this way it is still possible to develop ad hoc interfaces for Android or iOS platforms if needed.

\begin{figure}[hbtp]
\centering
\includegraphics[width= 0.95 \linewidth]{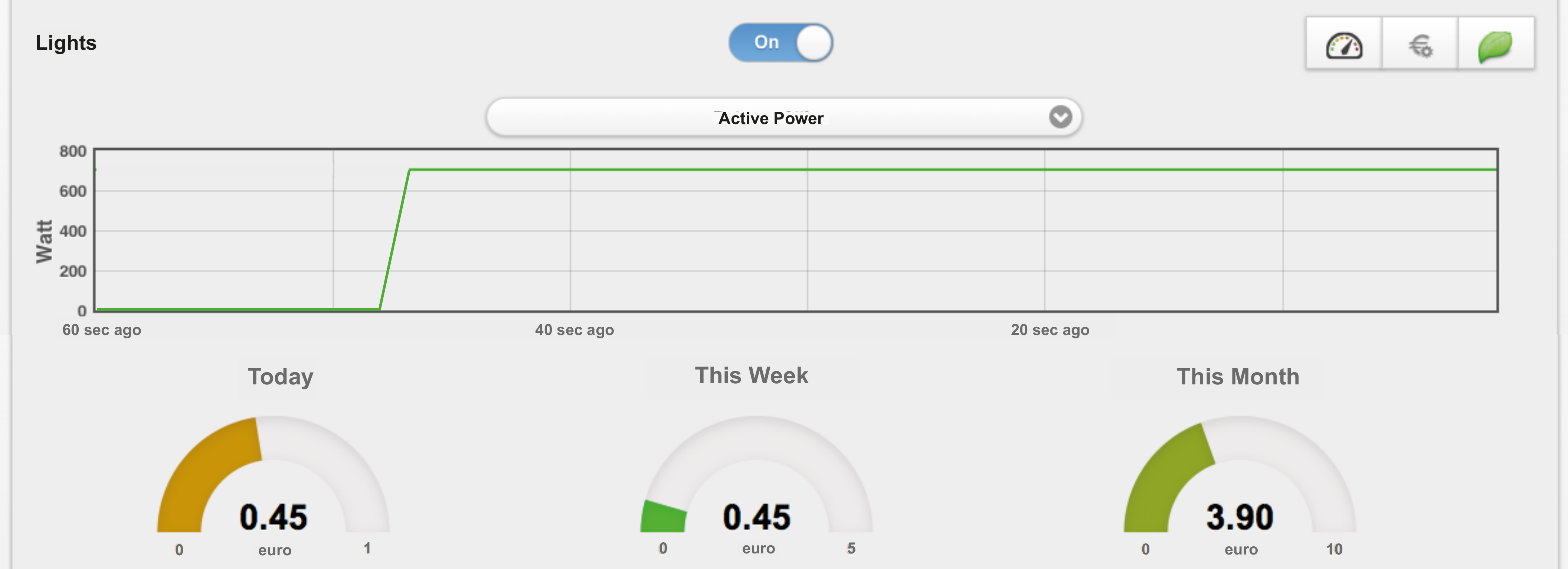}
\caption{Real-time power consumption}
\label{fig:realitme-screen}
\end{figure}

\begin{figure}[ht!]
     \begin{center}
		{\subfigcapskip = 0pt
		\subfigbottomskip = 0pt
        \subfigure[Historical data screen
        ]{%
            \label{fig:energy-screen}
            \includegraphics[width=0.52\columnwidth]{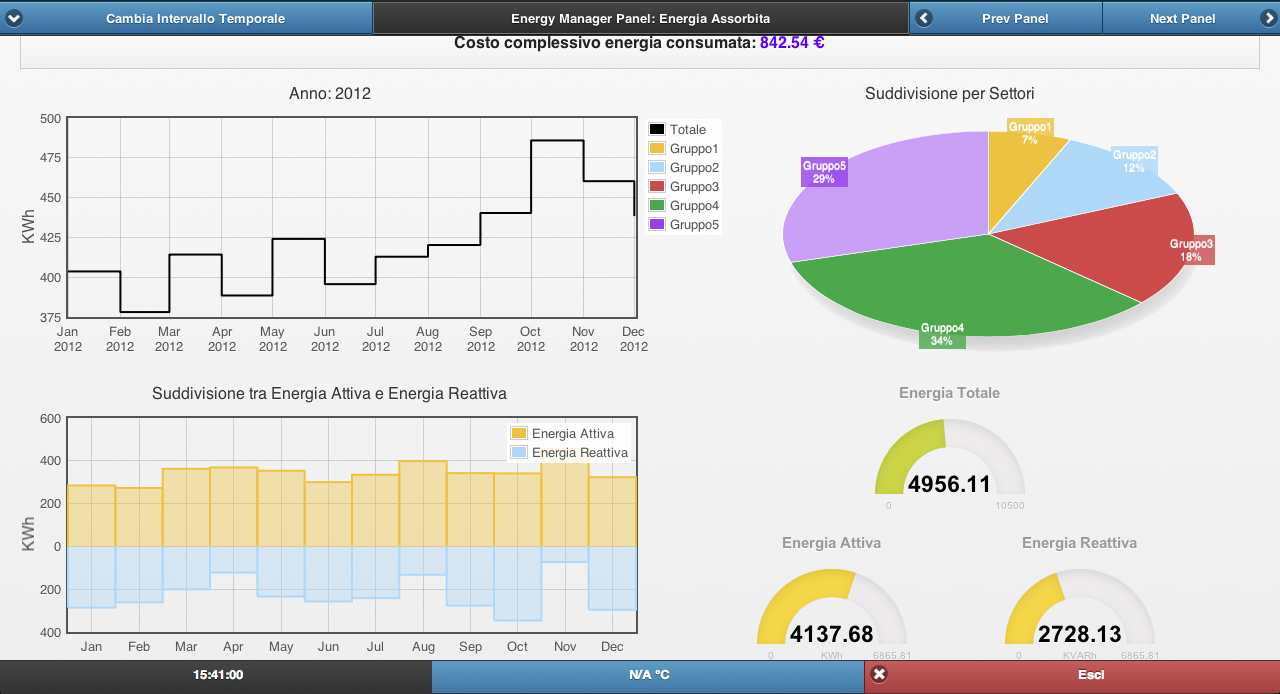}
        }%
        }
        \subfigure[Heat-map visualization
        ]{%
           \label{fig:heatmap-screen}
           \includegraphics[width=0.38\columnwidth]{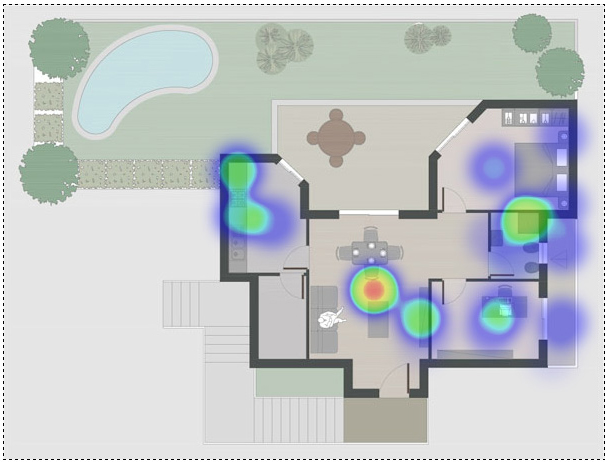}
        }\\ 
    \end{center}
    \caption{%
        Energy monitoring visualizations
     }%
   \label{fig:lights}
\end{figure}

By continuously monitoring the bus, the \obox{} is able to intercept power readings sent by the \ometer{}s and consequently is able to provide real-time feedbacks (Fig. \ref{fig:realitme-screen}) to the clients and store them for future use. 
Collected measurements can be retrieved later in a raw textual form or can be visualized in a graphical panel summarizing all the relevant information (cf. Fig. \ref{fig:energy-screen}), or in other visual forms (e.g., the heat map shown in Fig. \ref{fig:heatmap-screen}).

%% file: banks.tex
\section{Energy savings in offices}
\label{sec:banks}


The offices of an institution have been clustered on the basis of their extension, in order to identify three typical classes: small-sized offices (less than 300 sq.m.), medium-sized offices (extension between 300 sq.m. and 1000 sq.m.) and big-sized offices (greater than 1000 sq.m.). Then an office for each class has been identified (all the three offices in the same town, in order to avoid distortion effects on the outcomes of the study), and such three offices have been equipped with the \system~and monitored during August--October 2015.

\begin{figure}
\centering
\includegraphics[width=0.99\linewidth]{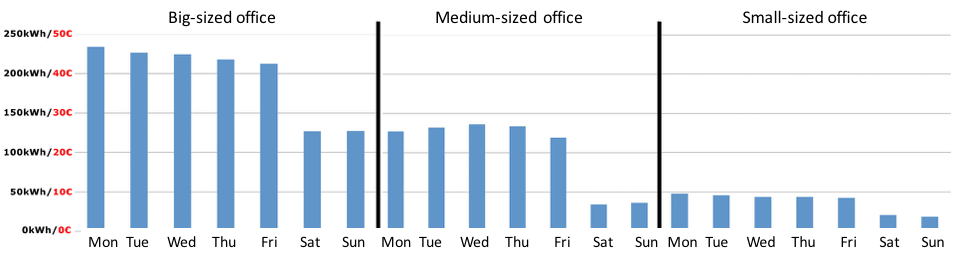}
\caption{The week consumption of the three offices}
\label{fig:weekConsumption}
\end{figure}

Fig. \ref{fig:weekConsumption} shows the average consumption of each office during a week, expressed in KWh and corresponding cost (assuming 0,20€/Wh as energy unit cost). Notably, the consumption is quite high during week-ends, but the medium-sized office is more efficient than the others (its consumption during week-ends is 25\% of the working days, whereas for the other cases is about 75\%). 

\begin{table}
	\caption{Baselines}
	\label{table:baseline}
	\centering	
\begin{tabular}{@{\hspace{0.3cm}} l @{\hspace{0.3cm}} c @{\hspace{0.3cm}}c @{\hspace{0.3cm}}c}
     \toprule
	 \textbf{} & \textbf{Big-sized} & \textbf{Medium-sized} & \textbf{Small-sized}\\
	 \midrule 
	 Measured minimal  & 4341 W & 905 W & 620 W \\ 
	 consumption \\
	 Average hourly  & 5552 W & 2710 W & 1017 W \\ 
	 consumption during\\
	 closing hours \\ \bottomrule
\end{tabular} 
\end{table}

Table \ref{table:baseline} shows the baseline for each office, i.e., the minimum value (in Watt) that has been recorded during the period of observation, and compares this with the average consumption during the nights. Notably, in an ideal case these two should be quite the same, and conversely they are quite different. Further analyzing such data (by decomposing the consumption over the different lines, the \system~allows this) we can discover that in the case of the big-sized office the consumption during nights might be reduced to 1165 W, for the medium-sized office to 867 W and for the small-sized office to 650 W. This can be achieved by switching off lights and other environment-related equipment (air-conditioning, etc.) -- which account for about 80\% of the consumption, and keeping on all the non-interruptible equipment (allarms, TV for security and control, etc.). The switching-off can be performed automatically through the same \system, which can act as actuation platform as well and configured through appropriate user-friendly configuration (the \system~allows defining rules through a graphical language based on Blockly\footnote{\url{https://developers.google.com/blockly/}}).

%% file: switch.tex
\section{Smart energy switch}
\label{sec:switch}



Nowadays, renewable energies\added{\cite{bull2001renewable}}, and photovoltaic (PV) in particular, have been indicated as the best solution for a major independence from the hydrocarbons. The possibility of independently producing energy for covering the consumptions of private producers resulted in a strong growth of the number of private PV plants. However, the introduction of such energy into the public grid can represent a problem in terms of sustainability that strongly compromises the PV advantages\added{\cite{carrasco2006power}}. In this sense, ``self consumption''~\cite{epia_position_paper}, intended as the capability of using the energy before it is reintroduced into the grid, can bring real benefits from the employment of PV, even for little private plants. In this section we illustrate the design and prototype implementation of a tool, called OEnergySwitch, which aims at maximizing the quantity of produced energy that is self-consumed in a domestic environment. 

%
\added{In general a solution that tries to maximize the self-consumption~\cite{carrasco2006power} indirectly reduces or avoids the grid-plant interactions.}
More in details, the \added{exceeding} energy produced by private PVs is sold to the provider,\added{\cite{Castillo-Cagigal2011}}; this technique is known as ``{\it net metering}'': \replaced{it has been the main-stream solution of the former PV systems, but in general it is inefficient from the point of view and of the energy balance and, often, of the saving.}{Its problem is that for the producer/customer the selling price is much lower than the buying one, so this mechanism is not convenient for the producer/customer and the saving is low.} Modern systems should try to use directly the energy, and to send to the grid just the exceeding one, thus the preferable approach is to maximize the immediate consumption of the energy produced by the plant. This technique is named ``self-consumption''. Clearly, the self-consumption is more convenient\replaced{, but}{; however}, at the state of the art, it is necessary to dimension the plant for providing a peak energy higher than the real need due to the impossibility of dividing the loads. The aim is therefore \added{to exploit the micro.-accounting capability of the \system} to design a tool that can split the total load into two sets $S1$ and $S2$ such that $S1$ contains the elements that maximize the quantity of self-consumed energy, while $S2$ is provided by the grid.

Our prototype operates in an house connected to the electric grid and to an off-grid photovoltaic small private plant\footnote{The dimension of the plant may range: if consumptions are high, the plant might produce up to 1 kW; otherwise, if the domestic consumption is lower, the total renewable production can be lower. We consider a range between 400 and 1000 W.}.
The energy switch works at outlet granularity: each single outlet can be fed using either the renewable energy or the classical one. This implies an higher capability of optimization; when the energy consumption of the appliances exceeds the production due to few watts, this approach allows to turn off the minimum energy quantity that makes the total need satisfiable by the PV. 

A working prototype implementing the energy switch has been built:  a small real hybrid PV system has been set; the inverter is an Opti-Solar SP Efecto 1000, that provides until 800 Watts of energy. The inverter gives the priority usage to the solar power. Then if it is not sufficient, it takes energy from the battery. As last, if the battery is discharged, it uses the grid's power. So it can work in UPS mode~\cite{karve2000three}. Moreover it  is equipped with a serial communication board based on RS232 protocol. The first characteristic provides a ``free'' protection system in case of wrong configuration, the second one provides an easy way for an informations exchange between the inverter and the micro-controller. 
The PV subsystem is completed by two panels, for a total maximum production of 220 Watts. This value is the DC power production, the effective AC outgoing from the inverter is lower, depending on the weather and the conversion lost. Since the inverter has a maximum applicable load and the environment is not fixed, a fuse should protect it from an excessive overload, in case of error.

The physical switching is implemented using relays. For the prototype we used 4 contact, 7 ampere, industrial relays. They are controlled through alternate current, so, if the prototype is composed by 3 relays, they are controlled by three \ometer~outs. The other three outs are used for feeding the loads. Practically, half of the gates of the \ometer~are delegated to loads feed, while the other half controls the relays, redirecting them to a source rather than another one. 
The micro-controller is an Olimex A-20 OLinuXino Micro, a dual core Cortex A7 1GHz frequency board with 1 GB RAM. The prototype developed is a little system composed by three outlets connected both to the inverter and to the grid. 

\begin{figure}[!h]
	\centering
	\subfigure[Complete prototipe, external view]
	{\includegraphics[width=0.4\columnwidth]{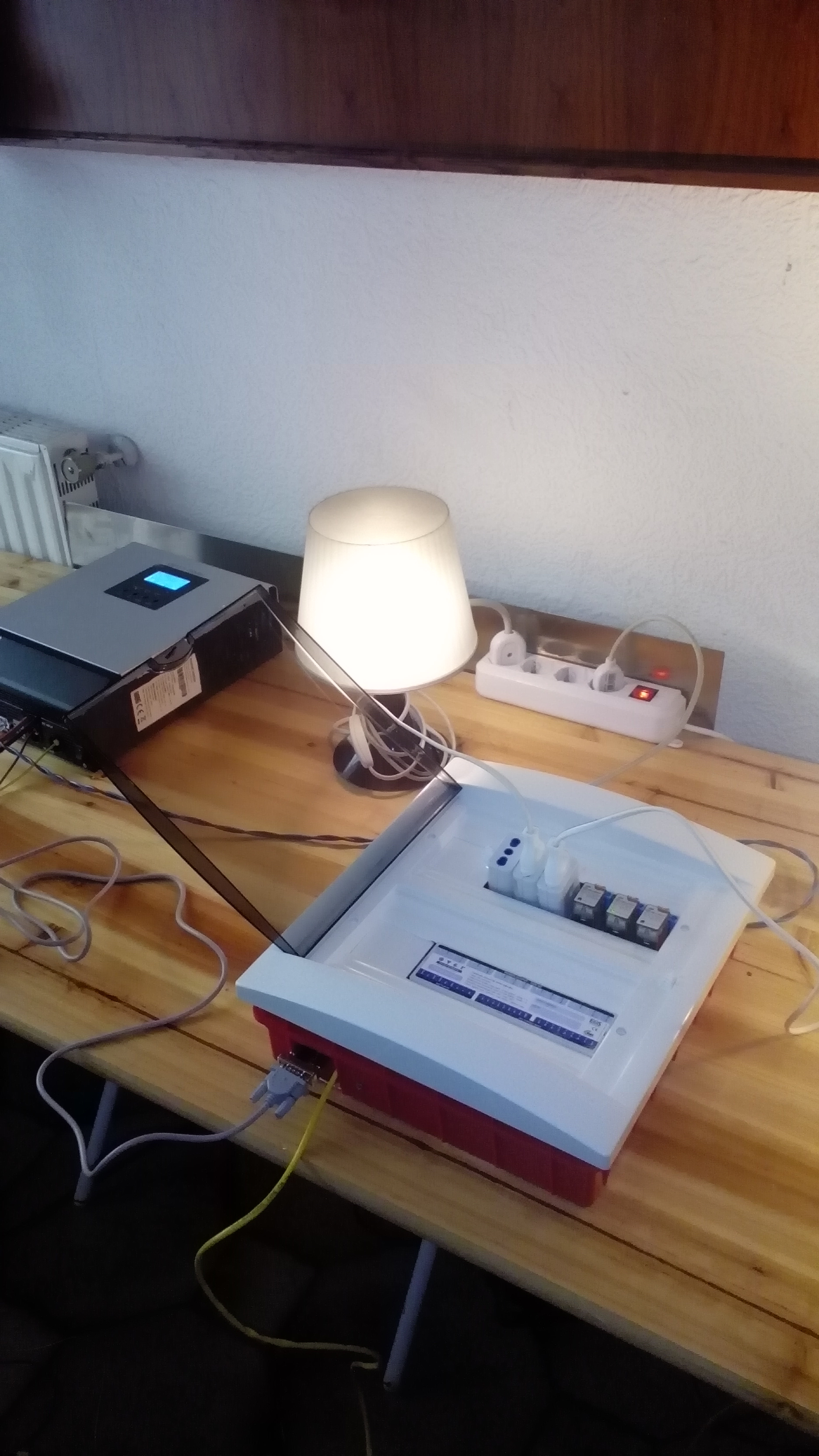}}
	\hspace{5mm}
	\subfigure[Energy Switch core, internal view]
	{\includegraphics[width=0.4\columnwidth]{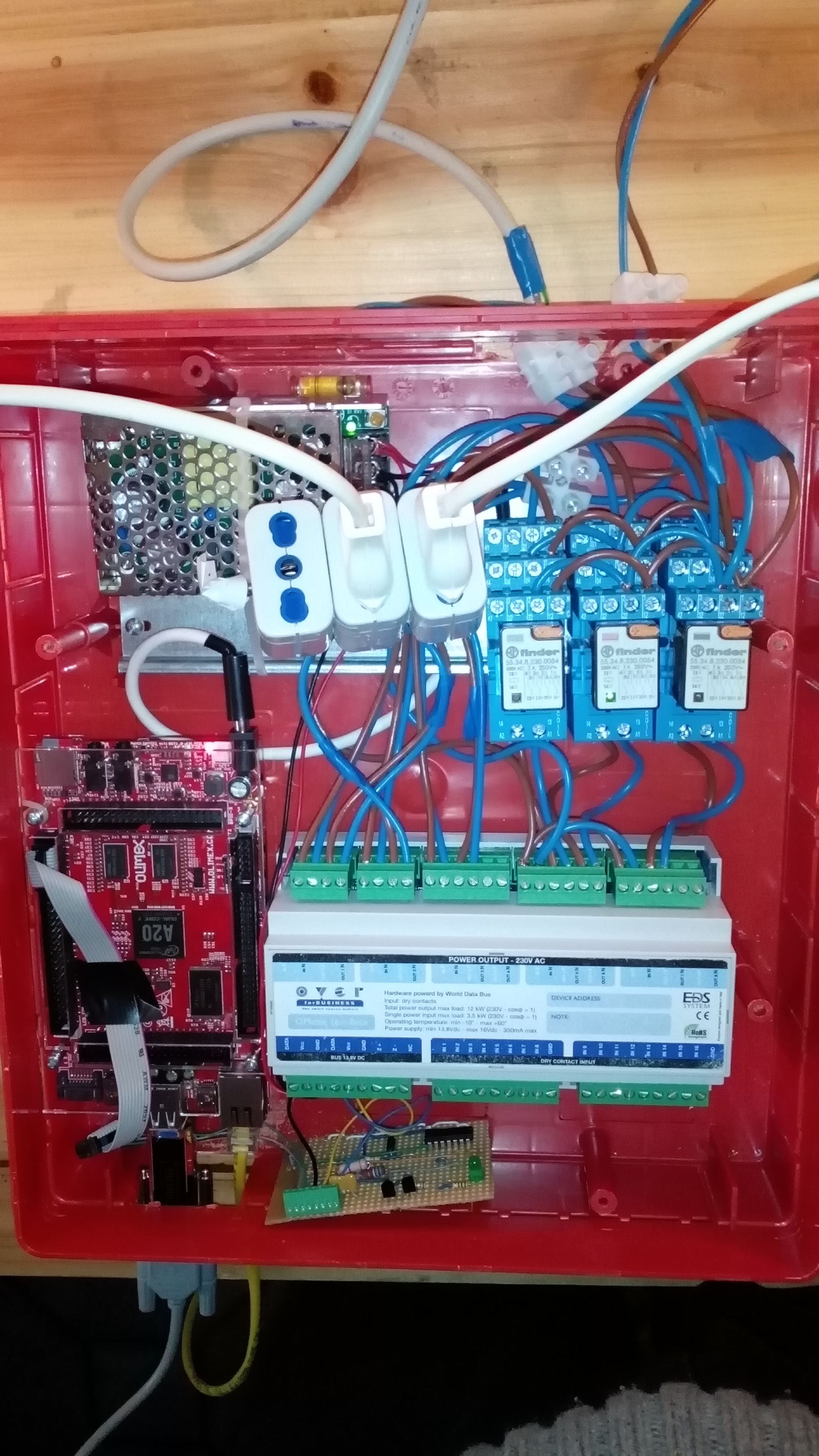}
		\label{e_view_a}}
	\caption{
		Two views of the prototype: Figure (a) represents the external view, with the inverter on the left and the core on the right. Figure (b) shows the different internal components: the Olimex, the OMeter and the relays block.   
		\label{internal_view_b}}
	\label{es_photo}
\end{figure}

The system computes the best outlet configuration and connect a subset of the total outlets to the domestic plant and lets the other to be fed by the grid energy.
The fine granularity energy monitoring capability allows a precise control on the outlets configuration. In particular, combining the information about the produced energy from the inverter and the data about the monitoring activity of the outlets, the system computes the solution of a knapsack problem for maximizing the direct consumption of the energy produced. The knapsack problem is modeled as follows:	\[
OP (i, w) = 
\left\{
\begin{array}{l l}
0 & \quad \text{if $i = 0$} \\ 
OP(i - 1 , w)  & \quad  \text{if $(w_i > w_k)$}         \\ 
max \left\lbrace OP(i-1, w) , v_i + OP (i-1, w-w_i)\right\rbrace & \quad \text{otherwise} 
\end{array}
\right.
\]
where $OP(i,w)$ is the maximum value subset of outlets $1, \dots,i $ with consumption limit $w$ (that is the capacity of the knapsack), $i$ is the $i$-th item, $w_i$ is its weight (apparent power absorbed) and $v_i$ is its value. So the limit is represented by the PV instantaneous production, and the power absorption constitutes the value and the weight of each outlet. The naive solution of the knapsack problem has an exponential computational cost. The implementation is based on a dynamic programming algorithm that can resolve the knapsack in a pseudo polynomial time. More in details each value is rounded to the nearest integer value: the incoming energy is rounded down, while each single consumption is rounded up. So for the total energy amount we have that $W_c = \lfloor W \rfloor $ and $wa_i = \lceil w_i \rceil$ where $W$ is the instantaneous total panels energy amount and $w_i$ is the instantaneous consumption of the $i$-th outlet. \deleted{However} This approach works fine if the values are read without any delay. In a real application, there is the need of considering the total information delay: from the inverter side, the delay can be considered negligible -- indeed the production changes are quite slow, and in a second the change is not substantial. The delay for the monitoring system is quite significant. Indeed let's call this delay $L_{monitor}$. We have that $L_{monitor} = L_{r} + L_{p} + L_{e}$ where $L_{r}$ is the delay related to the maximum reading frequency of the monitoring system, $L_{p}$ the propagation delay of the information from the station to the computing unit and $L_{e}$ is the elaboration time required by the computing unit. From an analysis of these elements, it appears that $L_p$ is very small. Also $L_e$ is quite short, since the hypothesis of low producing plant. $L_r$ instead is around one second. A second is a quite high value in the electrical context, so the effective energy situation at the time of the actuation of the computed configuration can be different and an energy lack can happen. In order to avoid that, the approach proposed is based on the study of the previous behavior of each outlet and on the history of the consumptions. In particular, time is divided in time slots and for each time slot the mean and the variance of the consumptions are computed and considered for taking a decision. For choosing the best approach both the time slots size and the different logics for the system have been tested.


We took into account the following switching logics: \emph{(i)} naive, \emph{(ii)} threshold on variance, \emph{(iii)} threshold on variance/mean ratio, \emph{(iv)} adaptive on variance, and \emph{(v)} adaptive on variance/mean ratio. All of them take into account a safety margin. A safety margin is a percentage of the incoming energy that is not considered as available for covering hypothetical errors and energy lacks. On the one hand, the bigger is the safety margin, the lower is the saving; on the other hand, the shorter is the margin, the higher is the error probability. The naive approach has been considered just as a comparison term. It does not add any information on the read data, so it is the most speculative approach and counts many errors. For the remaining approaches we can detect two groups: the first two approaches are static, while the others are dynamic.

A static approach implements a logic that excludes an outlet from the knapsack computation if the value of the considered metric overcomes a fixed threshold. Moreover, the percentage of the safety margin is fixed a priori. On the contrary, a dynamic (or  adaptive) approach implements a logic that excludes an outlet from the knapsack computation if the value of the considered metric overcomes a dynamic threshold. The safety margin is defined dynamically as well: smaller is the value of the considered metric, lower is the percentage of the safety margin.

\begin{figure}[!h]
	\centering
	\subfigure[Energy Saved]
	{\includegraphics[width=0.48\columnwidth]{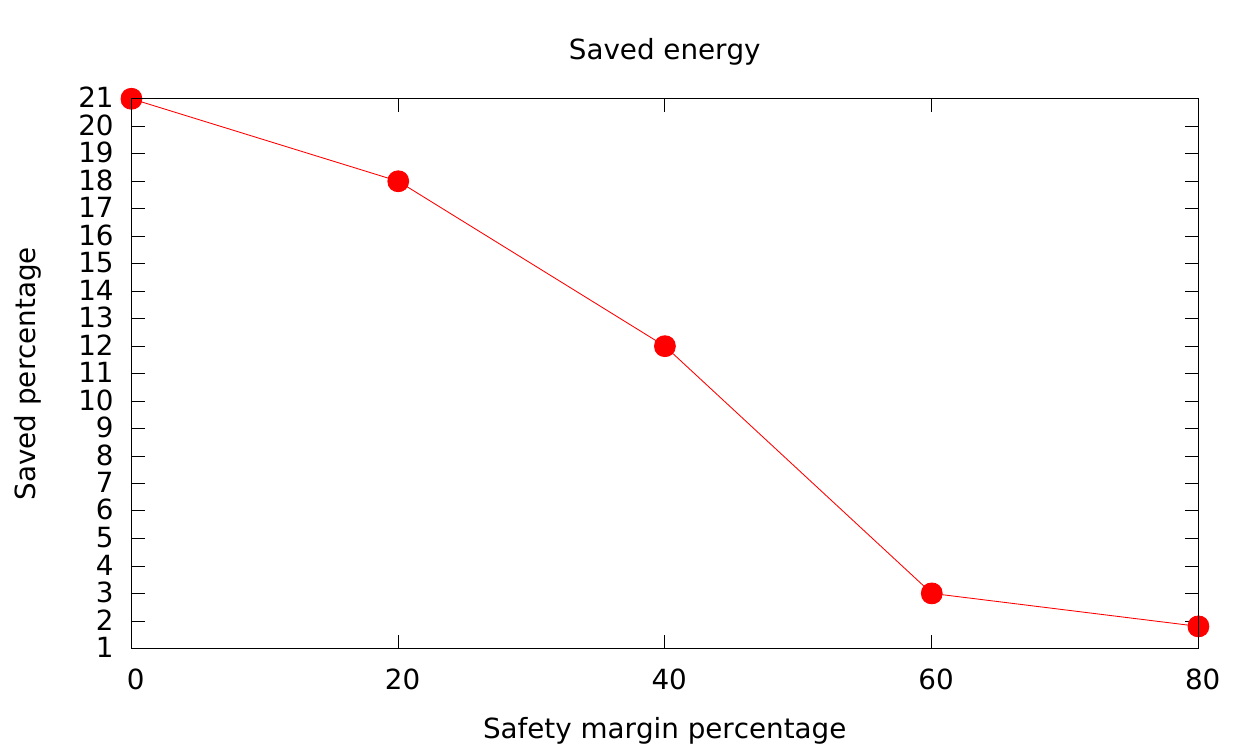}}
	\subfigure[Errors Occurred]
	{\includegraphics[width=0.48\columnwidth]{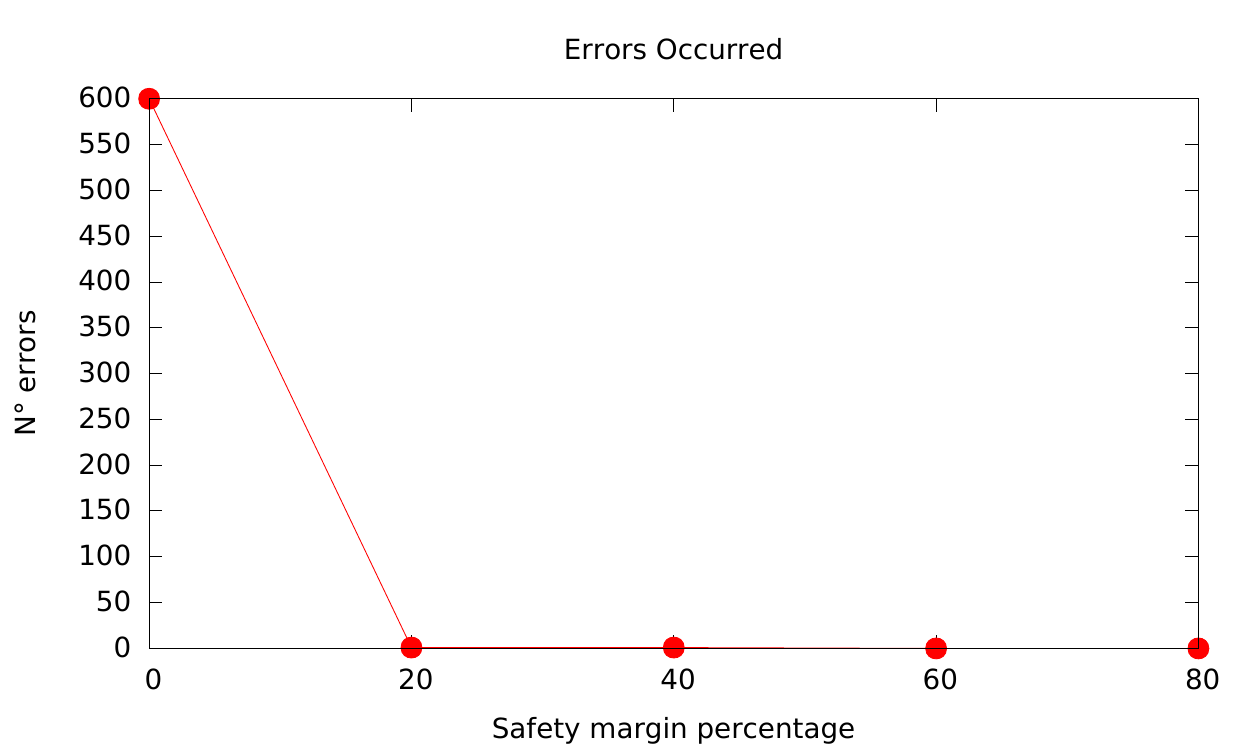}
		\label{tomv_simulated_approach48_a}}
	\caption{
		Threshold on $variance$ approach simulation results. The number of time slots considered is 48. The approach tries to minimize the risks when the benefits would be few. The best results are obtained with a 20\% safety margin.
		\label{tomv_simulated_approach48_b}}
	\label{tomv_simulated_approach48}
\end{figure}

Starting from the analysis of the naive approach results, a clear trend has been detected: the bigger the safety margin is, the lower is the number of total errors, but the energy saving slowly decreases. Even the introduction of a little margin causes a big reduction of the errors. However without a supplementary consideration the system appears not feasible in practical terms. On the contrary, if the supplementary metric analysis is introduced, the \replaced{performances}{performance} are quite good with a small safety margin of the 20\% (see Fig.~\ref{tomv_simulated_approach48}). The best results have been obtained with a time slot 30 minutes length: longer time slots give a less precise description of the consumption regularity, for shorter time slots can be difficult to catch the events and put the influence in the right time slot. However testing them in a real context the best performances have been reached with the adaptive approach using the variance. The saved energy has been the 17\% of the total, but the number of errors has been just two. This approach is very dynamic since it is not necessary to define the percentage of the safety margin as for the static approaches. {Hence, the results obtained suggest that the most important metric for describing the electric consumption is the variance, computed on 48 daily time slots. If the environment is well known and there are few changes in the system settings, the static approach can be adopted. It allows a greater save but it is weak against the changes. Otherwise the solution adaptive} is more general and does not require particular tuning. \deleted{However} Its saving is a little \added{bit} lower, so it is \replaced{a sound}{the right} \replaced{configuration also}{tuning} if the system setting is variable or not defined a priori. 

%% file: conclusions.tex
\section{Conclusions and Future Works}
\label{sec:end}

This dissemination paper has presented the \system~and a couple of applications built on top of it, notably how micro-accounting of energy allows to optimize consumption of offices and a new approach to the renewable energy management for small production plants in a domestic environments. In both cases, we demonstrated that the \system~can warranty a saving in terms of energy and money (about 20\% in both cases).

\begin{sloppypar}
The platform will be adopted in a recently funded H2020 project, namely GAIA, for accounting energy consumption and experiment new approaches to education of persons to save energy through gamification \replaced{approaches}{techniques}. Initial proof-of-concepts about this concepts have been realized, notably a smart space simulator to be used for giving immediate feedbacks to users (cf. \url{www.overtechnologies.com/casavirtuale}) and an Android app with challenges to be conducted by students in order to change their habits (cf. \url{https://play.google.com/store/apps/details?id=com.energyconsumption.diego\_cecchini.noenergywaste}). In future work we will report the outcome of such experimentations.
\end{sloppypar}

\noindent \paragraph{Acknowledgements.} This work has been partly supported by the H2020 EU project GAIA (grant 696029). The authors would like to thanks the many persons involved in the developments described in this paper, namely Diego Cecchini, Vincenzo Forte, Rodolfo Pallotta, Silvia Ruggiero and all the other developers at Over Technologies.

%% file: document.bbl
\begin{thebibliography}{10}

\bibitem{bull2001renewable}
S.R. Bull.
\newblock Renewable energy today and tomorrow.
\newblock {\em Proceedings of the IEEE}, 89(8):1216--1226, 2001.

\bibitem{carrasco2006power}
J.M. Carrasco, L.G. Franquelo, J.T. Bialasiewicz, E.
  Galv{\'a}n, R.C.~Portillo Guisado, M. Prats, J.I. Le{\'o}n, and
  N. Moreno-Alfonso.
\newblock Power-electronic systems for the grid integration of renewable energy
  sources: A survey.
\newblock {\em Industrial Electronics, IEEE Transactions on}, 53(4):1002--1016,
  2006.

\bibitem{carusoThesis}
M. Caruso.
\newblock {\em Service Ecologies, Energy Management and Accessibility in Smart
  Homes}.
\newblock PhD thesis, Dipartimento di Ingegneria Informatica, Automatica e
  Gestionale A. Ruberti, Sapienza, Università di Roma, 2015.

\bibitem{Castillo-Cagigal2011}
M. Castillo-Cagigal, E. Caama{\~n}o-Mart{\'\i}n, E.
  Matallanas, D. Masa-Bote, A. Guti{\'e}rrez, F. Monasterio-Huelin, and
  J. Jim{\'e}nez-Leube.
\newblock PV self-consumption optimization with storage and active DSM for the
  residential sector.
\newblock {\em Solar Energy}, 85(9):2338--2348, 2011.

\bibitem{ehrhardt2010advanced}
K. Ehrhardt-Martinez, K.A. Donnelly, S. Laitner, et~al.
\newblock Advanced metering initiatives and residential feedback programs: a
  meta-review for household electricity-saving opportunities.
\newblock American Council for an Energy-Efficient Economy Washington, DC,
  2010.

\bibitem{epia_position_paper}
EPIA – the European Photovoltaic Industry Association.
\newblock {\em Self Consumption of PV electricity}, July 2013.

\bibitem{han2011green}
J. Han, C.S. Choi, W.K. Park, and I. Lee.
\newblock Green home energy management system through comparison of energy
  usage between the same kinds of home appliances.
\newblock In {\em Consumer Electronics (ISCE), 2011 IEEE 15th International
  Symposium on}, pages 1--4. IEEE, 2011.

\bibitem{hart2008using}
D.G. Hart.
\newblock Using AmI to realize the smart grid.
\newblock In {\em Power and Energy Society General Meeting-Conversion and
  Delivery of Electrical Energy in the 21st Century, 2008 IEEE}, pages 1--2.
  IEEE, 2008.

\bibitem{karve2000three}
S. Karve.
\newblock Three of a kind [UPS topologies, IEC standard].
\newblock {\em IEEE Review}, 46(2):27--31, 2000.

\bibitem{masiello2010demand}
R. Masiello.
\newblock Demand response the other side of the curve [guest editorial].
\newblock {\em Power and Energy Magazine, IEEE}, 8(3):18--18, 2010.

\bibitem{ricciardi}
S.~Ricciardi, G. Santos-Boada, D. Careglio, F. Palmieri and U. Fiore, ``Evaluating energy savings in WoL-enabled networks of PCs,''
  in \emph{Proc. 2013 IEEE International Symposium on Industrial Electronics (ISIE)}.

\bibitem{rojchaya2009development}
S. Rojchaya and M. Konghirun.
\newblock Development of energy management and warning system for resident: An
  energy saving solution.
\newblock In {\em Electrical Engineering/Electronics, Computer,
  Telecommunications and Information Technology, 2009. ECTI-CON 2009. 6th
  International Conference on}, volume~1, pages 426--429. IEEE, 2009.
\end{thebibliography}
